# TEACHING PRECURSORS TO DATA SCIENCE IN INTRODUCTORY AND SECOND COURSES IN STATISTICS

Nicholas J. Horton[1], Benjamin S. Baumer[2] and Hadley Wickham[3]
1: Department of Mathematics, Amherst College, Amherst, MA USA
2: Department of Mathematics and Statistics, Smith College, Northampton, MA USA
3: Rice University and RStudio, USA
nhorton@amherst.edu

*Statistics students need to develop the capacity to make sense of the staggering amount of information collected in our increasingly data-centered world. Data science is an important part of modern statistics, but our introductory and second statistics courses often neglect this fact. This paper discusses ways to provide a practical foundation for students to learn to "compute with data" as defined by Nolan and Temple Lang (2010), as well as develop "data habits of mind" (Finzer, 2013). We describe how introductory and second courses can integrate two key precursors to data science: the use of reproducible analysis tools and access to large databases. By introducing students to commonplace tools for data management, visualization, and reproducible analysis in data science and applying these to real-world scenarios, we prepare them to think statistically in the era of big data.*

INTRODUCTION

In a world awash in data, there is a pressing need for people who are able to extract actionable information from data. As statistics educators, we play a key role in helping to prepare the next generation of statisticians and data scientists. Rachel Schutt succinctly summarized the challenges she faced as she moved into the workforce: "It was clear to me pretty quickly that the stuff I was working on at Google was different than anything I had learned at school." (Schutt and O'Neil, 2013). Moreover, the relevance of this disconnect between the increasingly complex analyses now demanded in industry and the instruction available in academia is not exclusive to Google. Gould (2010) cautioned that many of our courses do not answer the statistical questions students want to address.

Schutt and O'Neil (2014) describe a number of specific capacities needed to make sense of the data around us. These include data "grappling skills": visualization, knowledge of statistics, experience with forecasting and prediction along with sophisticated communication skills. Finzer (2013) argues for the importance of repeated experiences working with data to develop "data habits of mind". Inculcating our students with this capacity, distinct from specific technologies, may help them thrive in an era where data analysis techniques must respond in step with the collection of larger and more complex data. Nolan and Temple Lang (2010) argue that "the ability to express statistical computations is an essential skill," and that major changes to foster this capacity are needed in the statistics curriculum at the graduate and undergraduate levels.

In this paper, we discuss two precursors to data science that we believe should be incorporated into a range of introductory and second courses in statistics to ensure that students begin to develop the ability to frame and answer statistical questions with richer supporting data. The two precursors are the use of reproducible analysis tools and the integration of large databases into first and second stage courses. These "baby-steps" are a necessary but not sufficient way to engage students in data science. We detail our experiences with these changes and describe ways to integrate additional tools into the curriculum.

REPRODUCIBLE ANALYSIS AND R MARKDOWN

Recent efforts in statistics education have advocated for an increased use of computing in the statistics curriculum (American Statistical Association, 2000; Nolan and Temple Lang, 2010; American Statistical Association, 2013). At the same time, there has been a startling realization that many modern scientific findings cannot be reproduced (Nature, 2013). While there are many causes for irreproducibility of results (including issues of multiplicity and data-dredging), a copy-and-paste workflow—an artifact of antiquated user-interface design—makes reproducibility of statistical analysis more difficult, especially as data become increasingly complex and statistical methods become increasingly sophisticated (Baumer et al, 2014).

Reproducible analysis (Gentleman & Temple Lang, 2004; Xie, 2014) facilitates the ability of an analyst to conduct and present data analysis in a way that another person can understand and replicate (or that they can do themselves at a later point in time). This is the core of collaborative science. The emphasis on reproducibility can be seen as a necessary but not sufficient part of ensuring that students have capacity to "think with data." A natural environment to provide this mentoring is the first time most young scientists will encounter the formal principles of scientific inquiry and reproducibility: in introductory statistics.

R Markdown (Allaire et al, 2013) is a new technology that simplifies the creation of fully-reproducible statistical analysis. This system, in conjunction with the open-source RStudio environment (R Core Team 2013 and RStudio, 2013), enables students without prior knowledge of a markup language to combine statistical computing and written analysis in *one document*. Briefly, it transforms a well-annotated source file into a self-contained HTML file with embedded graphics, commands, and stylized text. By integrating the analysis, documentation, and interpretation, a structured workflow is provided to novice statisticians. This helps to minimize the pain of iterative analyses, and leaves behind a clearly marked set of instructions. Additional content in the form of text, lists, headers, tables, external images, and web links, etc. can surround the command chunks in a standard way.

Baumer et al (2014) describes experiences integrating R Markdown into introductory statistics courses in both small (Smith College) and large (Duke University) programs. With appropriate support mechanisms, introductory statistics students were receptive to its adoption. R Markdown was an improvement over the traditional copy-and-paste workflow. Students left the course equipped with the ability to undertake reproducible analysis, along with a clear sense for the advantages of this approach.

ACCESSING LARGE DATASETS WITHIN R

Nolan and Temple Lang (2010) also stress the importance of knowledge of information technologies, along with the ability to work with large datasets. Relational databases, first popularized in the 1970's, provide fast and efficient access to terabyte-sized datasets (Tahaghoghi and Williams, 2006). Connections between general purpose statistics packages such as R and database systems can be facilitated through use of SQL (structured query language). Such interfaces are attractive as they allow the exploration of large datasets that would be impractical to analyze using general purpose statistical packages (Ripley, 2001).

The use of SQL within R is straightforward once the database has been created. An add-on package (such as RMySQL or RPostgreSQL) must be installed and loaded, then a connection made to a local or remote database.

MOTIVATING EXAMPLE: AIRLINE DELAYS

These two precursors make it possible to realistically analyse a large dataset stored in a database in an introductory or second course. Students can use this to address questions that they find real and relevant (Gould, 2010), such as airline delays. It is not hard to find motivation for investigating patterns of flight delays. Ask students: have you ever been stuck in an airport because your flight was delayed or cancelled and wondered if you could have predicted it if you'd had more data? This dataset, which contains more than 150,000,000 observations corresponding to each commercial airline flight in the United States between 1987 and 2012, was utilized in the ASA Data Expo 2009 (Wickham, JCGS, 2011). The ASA Data Expo 2009 website (http://stat-computing.org/dataexpo/2009) provides full details regarding how to download the Expo data (1.6 gigabytes compressed, 12 gigabytes uncompressed through 2008), set up a database, add indexing, and then access it from within R and RStudio.

This opportunity to make a complex and interesting dataset accessible to students in introductory statistics is quite compelling. In the first course, this was introduced through use of the "Judging Airlines" model eliciting activity (MEA) documented by the CATALST Group (2009). This MEA requires no technology, but guides students to develop ideas regarding center and variability using small samples of data for pairs of airlines flying out of Chicago. Later in the course, students return to the informal "rule" they developed in an extension to determine whether to make the call about one airline being more reliable than the other. Their rule can be automated, and then carried out on a series of random samples from the flights from that city on that airline within that year. This allows them to see how often their rule picked an airline as being more reliable. Finally, students can summarize the population of all flights, as a way to better understand sampling variability. This process reflects the process followed by analysts working with big data: sampling is used to generate hypotheses that are then tested against the complete dataset.

The computation for the comparison of their informal "rule" and analyses of the distribution of the population values requires some coding (see examples at http://www.amherst.edu/~nhorton/icots2014). It would not be feasible to have students run these commands without some support. The provision of an instructor-provided R Markdown template (leveraging earlier work and demonstrating the power of the tools) allows R to be used as a tool in a small component of the course.

In a second course, more time is available to develop diverse statistical skills. This includes more sophisticated data management and manipulation, such as the calculation of the weekly count of flights over this period (reprising the display from Wickham (2009)) with additional years of data. This can be undertaken with a single SQL SELECT statement and some modest post-processing in R. Figure 1 displays the pattern, which has many interesting aspects.

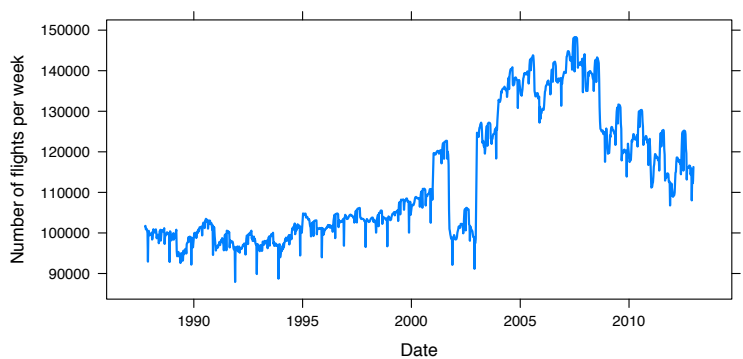

Figure 1: display of weekly counts of commercial flights in the United States over the period 1987 to 2012 (total n= 148,562,493). There is a strong seasonal pattern and clear impact of 9/11.

Other data wrangling and manipulation capacities can be introduced and developed using this example, including data joins/merges (since there are tables providing additional (meta)data about planes and airports). Linkage to other data scraped from the Internet (e.g. detailed weather information for a particular airport or details about individual planes) may allow other questions to be answered. Other approaches to analysis of big data in R (e.g. dplyr package) can also be introduced.

Use of a database to access this rich dataset helps to excite students about the power of statistics as well as introduce tools that can help energize the next generation of data scientists.

CONCLUSION

Nolan and Temple Lang argue that students need the facility to express statistical computations. In addition, there have been other calls for an increased use of computing in the statistics curriculum at the undergraduate level (American Statistical Association, 2000). In an era of increasingly big data, we agree that this is an imperative to develop in students, beginning with the introductory course. Some in the data science world argue that statistics is only relevant for "small data" and "traditional tools." We believe that the integration of these precursors to data science into our curricula—early and often—will help statisticians be part of the dialogue regarding Big Data and Big Questions (Davidian, 2013).

We concur that there are barriers and costs to the introduction of reproducible analysis tools and databases to our courses. Cobb (2007) argued that statistics courses are mired in teaching techniques developed by pre-computer-era statisticians to circumvent their lack of computational power. Further guidance and research results are needed to guide our work in this area. As Schutt and O'Neil (2013) caution, statistics could be viewed as obsolete if this challenge is not embraced.

Finzer (2013) noted that such changes are also needed before university level, and that the U.S. K-12 education system "does not provide meaningful learning experiences designed to develop understanding of data science concepts or a fluency with data science skills". He concludes that statistics educators—who generally understand data, have substantial expertise in computation, and have developed a variety of data habits of mind—are well-positioned to advocate for major changes in the training of future data scientists. We believe that the time to move forward in this manner is now, and believe that these two precursors provide a foundation for such efforts.

ACKNOWLEDGMENTS
This work was partially supported by Project MOSAIC, US NSF (DUE-0920350). We are indebted to Yihui Xie, J.J. Allaire, Jeffrey Horner, Vicent Marti, and Natacha Porte for their work on the knitr and markdown packages in R as well as the R Special Interest Group on Databases for the DBI interface.